# Using Overlap Weights to Address Extreme Propensity Scores in Estimating Restricted Mean Counterfactual Survival Times

**Authors:** Zhiqiang Cao[1], Lama Ghazi[2], Claudia Mastrogiacomo[3,4], Laura Forastiere[3,4], F. Perry Wilson[5], Fan Li[3,4]

**Correspondence to** Fan Li, PhD, Department of Biostatistics, Center for Methods in Implementation and Prevention Science, Yale School of Public Health, Suite 200, Room 229, 135 College Street, New Haven, Connecticut 06510 (email: fan.f.li@yale.edu)

**Author affiliations:**

[1]Department of Mathematics, College of Big Data and Internet, Shenzhen Technology University, Guangdong, China.

[2]Department of Epidemiology, School of Public Health, University of Alabama at Birmingham, Birmingham, Alabama, USA.

[3]Department of Biostatistics, Yale School of Public Health, New Haven, Connecticut, USA.

[4]Center for Methods in Implementation and Prevention Science, Yale School of Public Health, New Haven, Connecticut, USA.

[5]Clinical and Translational Research Accelerator, Department of Medicine, Yale School of Medicine, New Haven, Connecticut, USA

**Abbreviations:**

IPCW: inverse probability of censoring weighting

IPTW: inverse probability of treatment weighting

OW: overlap weighting

PS: propensity score

RMST: restricted mean survival time

RMCST: restricted mean counterfactual survival time

**Abstract**

While inverse probability of treatment weighting (IPTW) is a commonly used approach for treatment comparisons in observational data, the resulting estimates may be subject to bias and excessively large variance when there is lack of overlap in the propensity score distributions. By smoothly down-weighting the units with extreme propensity scores that are close to zero or one, overlap weighting (OW) can help mitigate the bias and variance issues associated with IPTW. Although theoretical and simulation results have supported the use of OW with continuous and binary outcomes, its performance with right-censored survival outcomes remains to be further investigated, especially when the target estimand is defined based on the restricted mean survival time (RMST)—a clinically meaningful summary measure free of the proportional hazards assumption. In this article, we combine propensity score weighting and inverse probability of censoring weighting to estimate the restricted mean counterfactual survival times, and propose computationally-efficient variance estimators when the propensity scores are estimated by logistic regression and the censoring process is estimated by Cox regression. We conduct simulations to compare the performance of IPTW, trimming, truncation and OW in terms of bias, variance, and 95% confidence interval coverage, under various degrees of overlap. Under moderate and weak overlap, we demonstrate via simulations the advantage of OW over IPTW, trimming and truncation methods, with respect to bias, variance, and coverage when estimating RMST.

**INTRODUCTION**

Observational data are frequently used to compare treatment-specific survival for drawing causal inference in epidemiological studies. Under the counterfactual outcomes framework (1), a common target estimand is the difference in average counterfactual outcomes, which represents the mean change had the entire population received a given treatment versus the standard care. In the presence of right-censoring that is inherent in survival data, the difference in average counterfactual survival times is inestimable, but its alternative, the restricted mean survival time (RMST) is estimable and easy to interpret (2-4). RMST measures the average survival time among the target population when the follow-up time is fixed at $L$, and corresponds to the area under the survival curve up to $L$. To address confounding in observational studies, a conventional approach to estimate the causal RMST estimand is through the multivariable Cox regression (5-6). However, if the proportional hazards assumption fails to hold, the estimated RMST may be subject to bias. Furthermore, under weak covariate overlap, direct outcome modeling is sensitive to model misspecification and can lead to biased inference due to over-extrapolation.

To avoid assumptions about the outcome model, we investigate propensity score weighting estimators for causal RMST estimands with a right-censored survival outcome. Since RMST is a summary of the survival function, one approach is to incorporate the inverse probability of treatment weighting (IPTW) with inverse probability censoring weighting (IPCW) into the Kaplan-Meier estimator (7) or Nelson-Aalen estimator (8)—nonparametric estimators for the survival function (9-10). A second approach is to combine IPTW with pseudo-observations (11-12). Each pseudo-observation is derived from a leave-one-out procedure and treated as an uncensored contribution to the target parameter, allowing for the standard method to proceed as if the outcomes were fully observed. Under both approaches, IPTW is constructed based on propensity scores (PSs, i.e., the probability of receiving treatment given pre-treatment covariates) and is needed to address confounding bias. With IPTW, each subject is weighted by the inverse of his/her probability of receiving the treatment he/she is assigned to given covariates (13-15), to balance the covariate distributions between the comparison groups. However, when there is weak overlap in the PS distributions such that the PSs for some units are close to 0 or 1 (or equivalently, PSs are extreme), the weights for these units can become excessively large; IPTW then tends to be influenced by these large weights, leading to biased and inefficient effect estimates (6, 16-17). Li et al. (18) and Cheng et al. (19) compared two methods that can improve upon IPTW under weak overlap, that is, trimming (16-17, 20) and overlap weighting (OW) (6, 18-19, 21). In addition to trimming, truncation is another useful technique to stabilize the inverse probability of treatment weights (22-23). In general, both trimming and truncation require the specification of a threshold, to which the causal effect estimates can be sensitive. By contrast, OW automatically assigns more weight to units with PSs close to 0.5 and less weight to units with extreme PSs close to 0 or 1, avoiding excessively large weights by construction and obviating the arbitrary specification of a threshold that may have an impact on the final estimates.

With a survival outcome, the advantage of OW over IPTW and trimming methods has been empirically demonstrated in Cheng et al. (19) for estimating the counterfactual survival functions. Parallel results when the target estimand is the difference in RMST, however, remain less available. Mao et al. (6) demonstrated efficiency gains in using OW in lieu of IPTW for several estimands (including the RMST estimand) through simulations, but did not provide theoretical results to support the optimality of OW. Zeng et al. (11) showed that OW leads to optimal efficiency when combining OW with pseudo-observations, but their study is mostly restricted to censoring completely at random and involves a computationally intensive procedure based on jackknifing. Different from these previous studies, we combine PS weighting (via the general balancing weights) and IPCW under the Nelson-Aalen estimator to target the RMST estimand We further provide a computationally-efficient variance estimator when the PS weights and censoring weights are estimated respectively by logistic regression and Cox regression. We show that, under some conditions, OW still leads to optimally efficient estimators for RMST estimands among the family of balancing weights. We conduct simulations to compare OW and IPTW, trimming and truncation methods, in terms of bias, variance, and 95% confidence interval coverage to elucidate the empirical performances of different estimators when the estimand is defined based on RMST.

## METHODS

### Notation and Estimands

Consider an observational study comparing two treatments with $n$ units and survival outcomes. For a given unit $(i = 1, ..., n)$, a set of pretreatment covariates $\boldsymbol{X}_i$ and treatment status $A_i$ (= $a$, with $a = 1$ for treatment and $a = 0$ for control) are observed. Denote $T_i$ as survival time of unit $i$, which is subject to the right censoring (e.g., due to drop-out or end of follow-up) at time $C_i$. Therefore, the observed survival time is $U_i = \min(T_i, C_i)$, and the censoring indicator is $\delta_i = I(T_i \leq C_i)$, where $I(\cdot)$ is an indicator function. We also represent $N_i(t) = I(U_i \leq t, \delta_i = 1)$ as the event indicator and $Y_i(t) = I(U_i \geq t)$ as the at-risk indicator at time $t$.

Under the counterfactual outcomes framework, we define $T_i^{(a)}$ as the counterfactual survival time and $C_i^{(a)}$ as the counterfactual censoring time of each unit if, possibly contrary to fact, s/he received treatment $A_i = a$. Within the balancing weights framework (13), the target population is defined as a tilted version of the observed population. Different choices of weighting scheme (e.g., IPTW, trimming, OW) corresponds to a specific target population. Here, we define the difference in restricted mean counterfactual survival time (RMCST) at restriction time $L$ ($L$ is smaller than the maximum follow-up time) in the target population as

$$\Delta_w(L) = \mu_w^{(1)}(L) - \mu_w^{(0)}(L) = E_w\{min(T^{(1)}, L)\} - E_w\{min(T^{(0)}, L)\}$$

$$= \int_0^L \{S_w^{(1)}(t) - S_w^{(0)}(t)\}dt, \qquad (1)$$

where the expectation is taken with respect to the target population, and $S_w^{(a)}(t)$ represents the counterfactual survival function (we use $w$ to index the target population which has a one-to-one mapping to the choice of balancing weight $w$). A formal exposition of this estimand and examples of balancing weights are provided in Web Appendix 1.

To estimate RMCST, the following standard assumptions for causal survival analysis with a point treatment (1, 19, 24) are adopted: Assumption 1—consistency and no interference (25), which assumes $T_i = T_i^{(a)}$ and $C_i = C_i^{(a)}$ for $A_i = a$; Assumption 2—exchangeability, which assumes away unmeasured confounders; Assumption 3—covariate-dependent censoring, which assumes that failure time is independent of censoring time given covariates in each group; Assumption 4—positivity, which assumes the conditional probability of treatment assignment is bounded away from 0 and 1, and the conditional survival probability of censoring is larger than 0 prior to maximum follow-up time.

**Propensity score balancing weights**

Mathematically, the PS is defined as the conditional probability of receiving treatment given the pre-treatment covariates $X_i$, that is, $e(X_i) = P(A_i = 1|X_i)$, and balancing the true PS is sufficient to balance all covariates. However, propensity scores are unknown in practice and must be estimated. We consider logistic regression to estimate the propensity scores based on the observed data, that is, $e(X_i, \beta) = 1/(1 + \exp(-\beta^T X_i))$. Then, the estimated PS is denoted as $\hat{e}(X_i) = e(X_i, \hat{\beta})$.

The balancing weight $\hat{w}_i$ of IPTW is defined as $1/\hat{e}(X_i)$ for treated units and $1/(1 - \hat{e}(X_i))$ for control units. The target population of IPTW is the combined treatment-control population represented by the study sample, and $\Delta_{IPTW}(L)$ measures difference in RMCST in this combined population. Because IPTW is prone to bias and excessive variance under weak overlap, we also consider *IPTW with trimming* and *IPTW with truncation* as two commonly-used modifications of IPTW to stabilize the estimation of causal effects. Two trimming approaches---symmetric trimming (16) and asymmetric trimming (17), as well as one truncation method (22-23) are considered. Under symmetric trimming, one first specifies a threshold $\alpha < 0.5$ and then excludes observations whose estimated PSs are outside of $[\alpha, 1 - \alpha]$. Crump et al. (16) suggested $\alpha = 0.1$ as a rule of thumb for non-censored outcomes. Under asymmetric trimming (17), one first excludes units whose PSs are outside of the common PS region formed by the treated and control units; then, one excludes all observations whose PSs are either below the $q$th quantile of the treated patients or above the $(1 - q)$th quantile of the control patients (26). Trimming removes units at the tail of the PS distribution to avoid the undue influence of extreme PSs, and the target population becomes the trimmed population (a subpopulation). Li et al. (18) recommended re-fitting the PS model based on the trimmed sample to ensure sample covariate balance and to improve the inference properties of IPTW with trimming estimators. Under truncation (22-23), we first obtain $p_q$ and $p_{1-q}$ as the $q$th, and $(1-q)$th quantiles of the PS distribution, and then set the lowest $100q\%$ and highest $100(1-q)\%$ of the PS distribution to be $p_q$ and $p_{1-q}$, respectively. Different from trimming, truncation maintains all units in the analysis but avoids the undue influence of extreme PSs by winsorizing the tail of the PS distribution; the intended target population under truncation remains the same as that under IPTW.

Finally, we consider the *overlap weighting* (OW) (21) approach, where $\widehat{w}_i = 1 - \hat{e}(X_i)$ for treated units and $\widehat{w}_i = \hat{e}(X_i)$ for control units. Under OW, the target population assigns larger weights to units with PSs close to 0.5 and smaller weights to units with extreme PSs close to 0 or 1 (because the tilting function $h(X_i) = e(X_i)\big(1 - e(X_i)\big)$ is defined as the product of PS and one minus PS; see Web Appendix 1). By doing so, OW corresponds to a target population with clinical equipoise whose treatment decisions are most uncertain; thus, the comparative effectiveness information for this target population is also highly relevant. Furthermore, OW has two attractive properties: (a) exact balance, that is when the propensity scores are estimated from logistic regression, the weighted covariate means are identical between the two treatment groups (21), and (b) minimum asymptotic variance, that is, under regularity conditions, OW leads to the most efficient estimator among the family of balancing weighting schemes for moment estimators. The target population of OW and $\Delta_{OW}(L)$ emphasizes units with the most overlap in their observed characteristics to achieve optimal internal validity (27).

**Combining PS weighting and IPCW for estimating RMCST**

We follow Robins and Finkelstein (7), Cheng et al. (19) to combine balancing weights and IPCW for addressing selection bias associated with right censoring and confounding bias associated with non-randomized treatment assignment. We model the survival distribution of the potential censoring time conditional on $X_i$ to estimate the censoring score, that is, $K_C^{(a)}(t, X_i) = P(C_i \geq t | X_i, A_i = a)$ for $a = 0,1$. There are several methods for estimating the censoring score including the Cox model (5), the additive hazard model (28) and parametric survival regressions (29). In what follows, we consider the Cox model (10, 30-31), i.e., $\lambda_C^{(a)}(t|X_i) = \lambda_0^{(a)}(t)\exp(\boldsymbol{\theta}_a^T X_i)$, where $\lambda_0^{(a)}(t)$ is the unspecified treatment-specific baseline hazard and $\boldsymbol{\theta}_a$ represents the regression coefficients. Estimators for $\boldsymbol{\theta}_a$ and $\Lambda_0^{(a)}(t) = \int_0^t \lambda_0^{(a)}(u)du$, denoted as $\widehat{\boldsymbol{\theta}}_a$ and $\widehat{\Lambda}_0^{(a)}(t)$, can be obtained by the maximum partial likelihood (32) and the Breslow method (33), respectively. For brevity, we define $\widehat{\Lambda}_C^{(a)}(t|X_i) = \widehat{\Lambda}_0^{(a)}(t)\exp(\widehat{\boldsymbol{\theta}}_a^T X_i)$ and $\widehat{K}_C^{(a)}(t, X_i) = \exp\left(-\widehat{\Lambda}_C^{(a)}(t|X_i)\right)$.

We notice that $E_w\{min(T^{(a)}, L)\} = \int_0^L \exp\left(-\Lambda_w^{(a)}(t)\right)dt$, for $a = 0,1$, where $\Lambda_w^{(a)}(t)$ is the cumulative hazard function of $T^{(a)}$ among the target population corresponding to weighting scheme $w$. Then an estimator for the RMCST estimand (1) at restriction time $L$ under weighting scheme $w$ is

$$\widehat{\Delta}_w(L) = \int_0^L \left\{\exp\left(-\widehat{\Lambda}_w^{(1)}(t)\right) - \exp\left(-\widehat{\Lambda}_w^{(0)}(t)\right)\right\}dt, \quad (2)$$

where the treatment-specific weighted Nelson-Aalen estimator (8, 10) is

$$\widehat{\Lambda}_w^{(a)}(t) = \int_0^t \frac{\sum_{i=1}^n I(A_i=a)\{\widehat{w}_i/\widehat{K}_C^{(a)}(u,X_i)\}dN_{ia}(u)}{\sum_{i=1}^n I(A_i=a)\{\widehat{w}_i/\widehat{K}_C^{(a)}(u,X_i)\}Y_{ia}(u)}, \quad (3)$$

and $N_{ia}(u) = I(A_i = a)N_i(u)$, $Y_{ia}(u) = I(A_i = a)Y_i(u)$ are the treatment-specific time-to-event indicator and the at-risk indicator at time $u$, respectively. In both the numerator and denominator of the integrand, the combined weight attached to each observation is given by $\widehat{w}_i/\widehat{K}_C^{(a)}(u, \boldsymbol{X}_i)$. The denominator of the integrand serves as a normalization factor to ensure that the normalized weights among the risk set at time $u$ sum to one for each treatment $a$. Therefore, estimator (3) is a Hajek-type estimator (rather than a Horvitz-Thompson-type estimator) similar to that defined in Zhou et al. (34). The normalization serves the purpose of stabilizing the weights with a potential efficiency gain in the estimator. Furthermore, the denominator of the integrand of $\widehat{\Lambda}_w^{(a)}(t)$ in (3) is a component of the type II estimator for the counterfactual survival function considered in Cheng et al. (19). However, the survival function estimator in Cheng et al. (19) is not guaranteed to produce monotonically non-decreasing survival functions, whereas one advantage of survival function estimator $\exp\left(-\widehat{\Lambda}_w^{(a)}(t)\right)$ is that it is by construction monotonically non-decreasing in follow-up time $t$; hence the counterfactual RMCST become monotone in $t$.

In Web Appendix 2, we prove that under assumptions 1-4, $\widehat{\Lambda}_w^{(a)}(t)$ is pointwise consistent to $\Lambda_w^{(a)}(t)$ under any PS balancing weight scheme, and establish the consistency of $\widehat{\Delta}_w(L)$ for finite $L$. We also show that, under regularity conditions, OW achieves the smallest asymptotic variance for estimating the difference in RMCST at each restriction time, completing the theoretical justification for using OW with the RMCST estimand.

Under the special case where the PS is estimated by a logistic regression and the censoring score is estimated by Cox regression, we develop a closed-form variance estimator for computationally efficient inference in Web Appendix 4. The variance estimator takes a simple form of

$\frac{1}{n^2}\sum_{i=1}^{n}(\hat{I}_{(\Delta,\beta),i} + \hat{I}_{(\Delta,\theta),i})^2$, where $\hat{I}_{(\Delta,\beta),i}$ and $\hat{I}_{(\Delta,\theta),i}$ quantify the impact on variability due to estimating the unknown PS and censoring baseline hazard. An interesting finding is that there is no impact on variance due to estimating the censoring score regression coefficients when the censoring score is estimated by Cox regression. Beyond this special case, bootstrapping is a useful and represents more general approach to quantify the uncertainty of $\widehat{\Delta}_w(L)$ and will be recommended.

**SIMULATION STUDIES**

*Simulation Design*

We conduct simulations to compare the performances of OW, IPTW, trimming and truncation in estimating the RMCST estimand under varying degrees of overlap between treatment groups. We follow Cheng et al. (19) and consider six covariates, i.e., $\boldsymbol{X} = (X_1, X_2, X_3, X_4, X_5, X_6)^T$. The first three covariates, $X_1 - X_3$, follow a multivariate normal distribution with mean zero, unit marginal variance, and a pairwise correlation coefficient of 0.5; the remaining covariates, $X_4 - X_6$, are independently generated from Bernoulli(0.5). The true PS is specified by a logistic model:

$$\text{logit}(e(\boldsymbol{X})) = \beta_0 + \beta_1 X_1 + \beta_2 X_2 + \beta_3 X_3 + \beta_4 X_4 + \beta_5 X_5 + \beta_6 X_6,$$

where $(\beta_1, \beta_2, \beta_3, \beta_4, \beta_5, \beta_6)^T = (0.15\gamma, 0.3\gamma, 0.3\gamma, -0.2\gamma, -0.25\gamma, -0.25\gamma)^T$. We vary $\gamma \in \{1,3,5\}$ to represent different degrees of overlap between groups; these three choices of $\gamma$ lead to strong, moderate, and weak overlap, respectively. The intercept is chosen such that 50% of units are assigned to each group. Visualization of the PS distributions under the three levels of covariate overlap are presented in Figure 1. We simulate counterfactual survival outcomes $T^{(a)}$ from an exponential distribution:

$$P(T^{(a)} > t|X) = \exp(-t\exp(m^{(a)}(X))), \quad a = 0,1,$$

where $m^{(1)}(X) = -1 + 0.4X_1 + 0.2X_2 + 0.1X_3 - 0.1X_4 - 0.2X_5 - 0.3X_6 + 2\text{logit}(e(X))$ under the treatment condition and $m^{(0)}(X) = -1.4 - 0.2X_2 - 0.3X_3 - 0.5X_4 - 0.6X_5 - 0.7X_6 - \text{logit}(e(X))$ under the control condition. We include the true PS in $m^{(a)}(X)$ to strengthen the correlation between the PS and the survival function. We assume there is no causal effect on censoring such that $C^1 = C^0 = C$, and generate $C$ from an exponential distribution $P(C \geq t|X) = \exp(-t\exp(\eta + \theta_a^T X))$, with $\theta_a = \theta = (-0.3, 0.5, 0.5, 0.2, -0.4, -0.5)^T$ and $\eta = -1.6$; this results in a marginal censoring rate around 50%. The observed survival time and censoring indicator are created by $U = \min(T, C)$ and $\delta = I(T \leq C)$, respectively.

For each scenario, we simulated 1000 datasets with sample size of $n = 1000$. For each dataset, we focus on estimating $\mu_w^{(1)}(L) = E_w\{\min(T^{(1)}, L)\}$, $\mu_w^{(0)}(L) = E_w\{\min(T^{(0)}, L)\}$ and $\Delta_w(L) = \mu_w^{(1)}(L) - \mu_w^{(0)}(L)$ at 3 pre-specified restriction time $L \in \{2,5,10\}$, and compare OW, IPTW, IPTW with PS trimming (abbreviated as trimming) and IPTW with PS truncation (abbreviated as truncation). We consider $\alpha = 0.05, 0.10, 0.15$ for symmetric trimming, $q = 0, 0.01, 0.05$ for asymmetric trimming; we consider $q = 0.025, 0.05, 0.10$ for truncation following previous simulations (16, 18-19, 23). Table 1 summarizes true values of estimands under each weighting scheme (obtained from a "super-population" by Monte Carlo methods), at each level of covariate overlap and restriction time; under weak overlap, the true RMCST estimands appear to be more different across different weighting schemes at a larger restriction time. For each method, we calculate bias, relative efficiency (Monte Carlo variance of IPTW divided by the variance for each weighting estimator), and 95% confidence interval coverage for each estimator. The reproducible R code for executing the simulations is available at:

https://github.com/Zhiqiangcao/Rcode_OW_RMST.

*Simulation results*

Table 2 summarizes the bias for each estimator with increasingly strong tails in the PS distributions. While OW, symmetric trimming, and asymmetric trimming with $q > 0$ have small bias, IPTW, asymmetric trimming with $q = 0$ and truncation can lead to noticeable bias under weak overlap ($\gamma = 5$). Furthermore, the asymmetric trimming with $q = 0$ and the truncation with $q \geq 0.05$ can exhibit bias even under moderate overlap ($\gamma = 3$). In addition, the bias of truncation increases with $q$. In general, OW minimizes the bias regardless of covariate overlap.

Table 3 presents the relative efficiency results. Throughout, OW is more efficient than IPTW, trimming and truncation methods for estimating RMCST estimands, and its efficiency advantage becomes more obvious when the degree of overlap decreases (e.g., 7 times more efficient compared to IPTW when $\gamma = 5$). Overall, symmetric trimming is more efficient than IPTW, asymmetric trimming and truncation, but less efficient than OW. Under weak overlap ($\gamma = 5$), asymmetric trimming with $q = 0$ can be even less efficient than IPTW and truncation, but setting $q > 0$ leads to a more efficient estimator compared to IPTW and truncation. Truncation is more efficient than IPTW in general, and its efficiency increases with the value of $q$.

In Table 4, we present the 95% confidence interval coverage based on the proposed closed-form variance estimator, when the propensity score and censoring score are estimated by logistic regression and Cox regression, respectively. OW and symmetric trimming with $\alpha =$ 0.15 provide nominal coverage under all three degrees of overlap, suggesting the adequacy of the proposed variance estimator. Coverage rates of the IPTW, asymmetric trimming approach with $q = 0$ and three choices of truncation can be far below nominal under moderate ($\gamma = 3$) and weak overlap ($\gamma = 5$). Symmetric trimming with two choices of $\alpha$ (i.e., 0.05 and 0.1) and asymmetric trimming approach with $q = 0.05$ correspond to slightly under-coverage with a large restriction time (e.g., $L = 10$). The coverage rate of IPTW quickly deteriorates when the degree of overlap decreases.

As a sensitivity check, we repeated the above simulation by omitting the PS term in generating the potential survival outcomes. That is, we consider $m^{(1)}(X) = -1 + 0.4X_1 + 0.2X_2 + 0.1X_3 - 0.1X_4 - 0.2X_5 - 0.3X_6$ and $m^{(0)}(X) = -1.4 - 0.2X_2 - 0.3X_3 - 0.5X_4 - 0.6X_5 - 0.7X_6$ in $P(T^{(a)} > t|X)$, while keeping all other simulation parameters unchanged. The results on bias, relative efficiency and interval coverage are qualitatively similar to our main simulations (Web Tables 5-8).

We also repeated our main simulations with a total sample size of 250 to further investigate the performance of OW against IPTW in smaller samples, and additionally to assess the performance of our closed-form variance estimators against bootstrapping. We consider 200 bootstrap replicates to create percentile-based confidence intervals when assessing coverage, and the results are presented in Table 5. The patterns on bias and relative efficiency are similar to our main simulation results, but the magnitude of relative efficiency for OW over IPTW appears smaller (the highest relative efficiency is 2 under lack of overlap). The interval coverage for OW is consistently closer to the nominal level compared to IPTW under moderate ($\gamma = 3$) and weak overlap ($\gamma = 5$), regardless of the use of the closed-form variance approach or bootstrapping. Bootstrapping can slightly improve the coverage over the closed-form variance approach with $n = 250$ (more substantial improvement for IPTW than for OW), likely because bootstrapping directly incorporates the uncertainty for estimating the censoring scores in small samples (while we have shown that this part does not contribute to the derivation of the asymptotic variance). Finally, we carried out a simulation study to compare different weighting estimators when the censoring model is incorrectly specified by Cox regression. The results are reported in Web Tables 9-11 in Web Appendix 8, and indicate that OW appears to be more robust to such misspecification than IPTW. This finding is similar to Zhou et al. (34) who have investigated similar comparisons with continuous outcomes when the propensity score model is incorrectly specified.

To summarize, the simulation results confirm that OW produces frequently minimal bias, optimal efficiency, and nominal coverage under different levels of overlap for estimating RMCST estimands. Although symmetric trimming, asymmetric trimming and truncation can improve over standard IPTW under weak overlap, they are still less efficient than OW and their performances may critically depend on the trimming or truncation threshold.

**DISCUSSION**

In this article, we combined PS weighting and IPCW to estimate RMCST estimands with observational data. In addition, we developed closed-form variance estimators when the PS is estimated by logistic regression and the censoring score is estimated by Cox regression, which can be considered as a computationally convenient alternative to bootstrapping in large samples.

However, in more general settings when the PS and censoring score are estimated by other models, bootstrapping remains useful and is recommended. In simulations, we show that OW leads to causal effect estimates with minimal bias and the highest efficiency, and consistently outperforms IPTW, trimming and truncation under moderate and weak overlap. Under strong overlap, OW performs similarly to and may occasionally be slightly more efficient than IPTW. Although lack of overlap is the motivation for developing OW, it should not be the sole reason to consider OW. The similarity in performance between IPTW and OW under strong overlap suggests that OW may be used more generally; Web Appendix 7 provides an illustrative analysis of an observational data with reasonable overlap to demonstrate that OW leads to narrower confidence intervals for RMCST compared to IPTW.

While the properties of different PS weighting methods cannot be established by one simulation study, the observations presented here are not unique. There has been an increasing amount of recent evidence to demonstrate the improved performance of OW over IPTW in observational studies with multiple treatments (35), survival outcomes (11, 19), subgroup analyses (36), and under model misspecification (34). Our study adds to that body of evidence with a focus on estimating causal effects on the unit of survival time scale. In practice, a common concern to switch from IPTW to OW is that OW corresponds to a different target population from the combined population (targeted by IPTW). However, a likely more critical question to ask is whether the combined population in the original study is truly reflective of the target population of scientific interest. The presence of extreme PSs in the study sample already indicates that certain units are almost always receiving the treatment (or control) and are considered outliers. It is then debatable whether comparative effectiveness evidence is of primary interest for such outlying units. OW addresses this issue from a design perspective, targets a subpopulation with the most overlap in their observed characteristics between treatments, and emphasizes units with clinical equipoise. This target population can be interpreted numerically (by summarizing weighted baseline covariates in a standard Table 1) and consists of units whose treatment decision is the most uncertain, and for whom the causal evidence is usually most needed. Under strong overlap, the true estimands of IPTW and OW have negligible differences and under a randomized clinical trial, their estimands even coincide (37-38). When covariates in treatment and control groups become increasingly separate, differences in estimands will increase. Finally, we acknowledge that this work considers a parametric, low-dimensional setting where it is possible to estimate logistic PS model to adequately balance covariates. In the presence of high-dimensional covariates (some of which may be strong predictors of the treatment and thus lead to separation issues), regularization methods would be required, but modifications to conventional regularization methods are usually needed to minimize the bias in the final causal effect estimates. For instance, the outcome-adaptive lasso (39) has been proposed for confounder selection with high-dimensional potential confounders; this approach regularizes the logistic regression with a $L^1$ penalty, which depends on the strength of correlation between each potential confounder and the outcome (estimated by a preliminary outcome model fit). Furthermore, hybridizing PS weighting and regularization methods such as combining OW with lasso (40-41) have been investigated in the context of causal subgroup analyses, although the prior focus has not been on RMCST. It may be worthwhile to extend those prior efforts to address high-dimensional confounders when the target estimand is RMCST.

To facilitate the application of the proposed methods, we provide sample R code and a step-by-step tutorial with simulated data in Web Appendix 5. Analyses of two observational studies are presented in Web Appendices 6 and 7 to illustrate the application of IPTW and OW for estimating RMCST.

ACKNOWLEDGEMENT

This work was supported in part by National Institutes of Health grants P30DK079310, R01HS027626 and R01DK113191. The contents of this article are solely the responsibility of the authors and do not necessarily represent the views of National Institutes of Health. The hypertension study data set are not currently available due to ethical and privacy restrictions. The authors thank Cheng Chao for help with discussion of the methods and simulations.

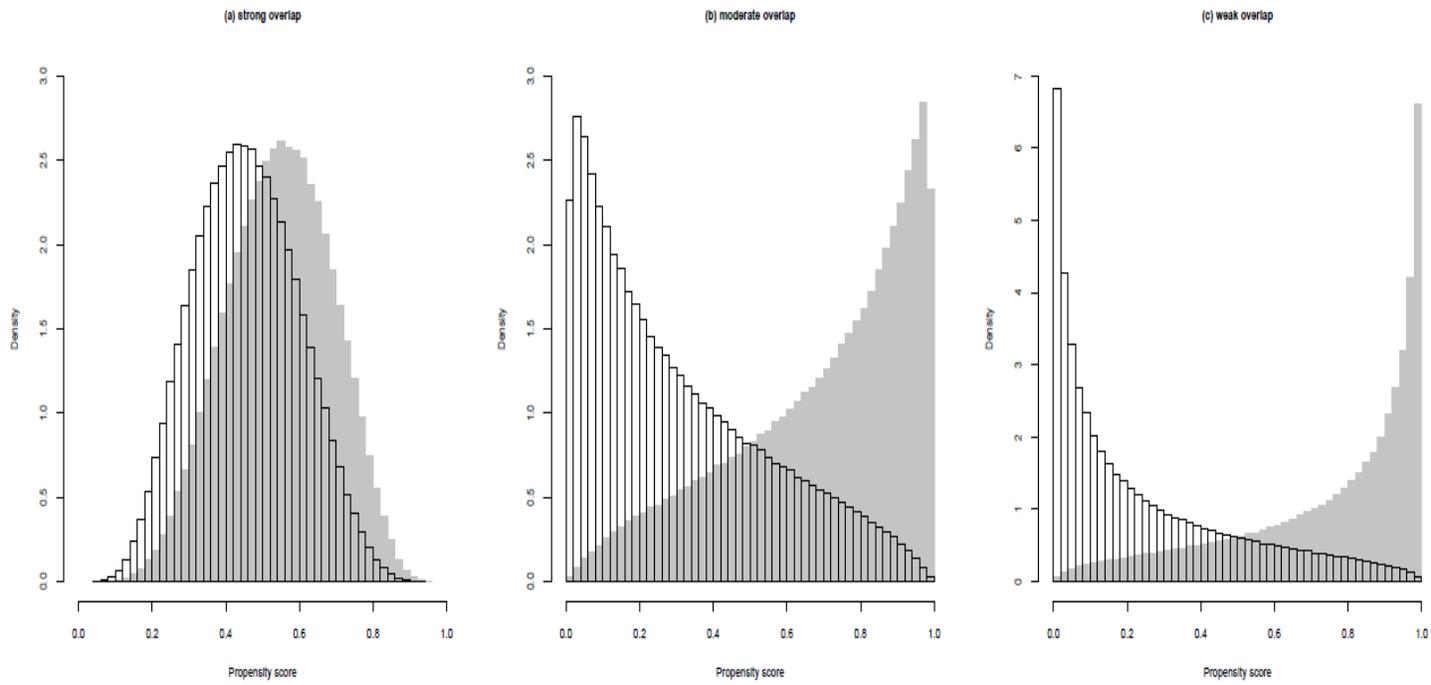

**Figure 1:** Distributions of true propensity scores with three levels of covariates overlap in simulation design, where the grey bars denote the treated group and the white bars represent the control group.

**Table 1.** True Values of $\mu_w^{(1)}(L)$, $\mu_w^{(0)}(L)$ and $\Delta_w(L)$ for Three RMCST at Strong, Moderate and Weak Levels of Covariate Overlap (Simulation Study).[a]

| Estimator | $\gamma$ | $\mu_w^{(1)}(L)$ | | | $\mu_w^{(0)}(L)$ | | | $\Delta_w(L)$ | | |
|---|---|---|---|---|---|---|---|---|---|---|
| | | $L=2$ | $L=5$ | $L=10$ | $L=2$ | $L=5$ | $L=10$ | $L=2$ | $L=5$ | $L=10$ |
| Overlap Weighting | | | | | | | | | | |
| | 1 | 1.026 | 1.497 | 1.741 | 1.854 | 4.183 | 7.207 | -0.828 | -2.687 | -5.465 |
| | 3 | 1.025 | 1.517 | 1.787 | 1.852 | 4.176 | 7.191 | -0.827 | -2.659 | -5.404 |
| | 5 | 1.026 | 1.505 | 1.752 | 1.853 | 4.178 | 7.193 | -0.826 | -2.673 | -5.442 |
| Inverse Probability Treatment Weighting | | | | | | | | | | |
| No Trimming | | | | | | | | | | |
| | 1 | 1.024 | 1.518 | 1.797 | 1.853 | 4.179 | 7.198 | -0.829 | -2.661 | -5.402 |
| | 3 | 1.016 | 1.636 | 2.099 | 1.845 | 4.147 | 7.136 | -0.829 | -2.511 | -5.038 |
| | 5 | 1.012 | 1.693 | 2.245 | 1.841 | 4.132 | 7.107 | -0.829 | -2.438 | -4.862 |
| Symmetric Trimming | | | | | | | | | | |
| $a = 0.05$ | 1 | 1.024 | 1.518 | 1.797 | 1.853 | 4.179 | 7.198 | -0.829 | -2.661 | -5.402 |
| | 3 | 1.020 | 1.574 | 1.923 | 1.850 | 4.165 | 7.170 | -0.829 | -2.591 | -5.247 |
| | 5 | 1.023 | 1.545 | 1.839 | 1.851 | 4.170 | 7.180 | -0.828 | -2.626 | -5.341 |
| $a = 0.1$ | 1 | 1.024 | 1.518 | 1.795 | 1.853 | 4.179 | 7.199 | -0.829 | -2.661 | -5.404 |
| | 3 | 1.024 | 1.521 | 1.786 | 1.853 | 4.177 | 7.192 | -0.828 | -2.656 | -5.406 |
| | 5 | 1.027 | 1.489 | 1.704 | 1.854 | 4.182 | 7.202 | -0.827 | -2.694 | -5.498 |
| $a = 0.15$ | 1 | 1.025 | 1.513 | 1.782 | 1.853 | 4.181 | 7.201 | -0.829 | -2.667 | -5.419 |
| | 3 | 1.028 | 1.477 | 1.680 | 1.855 | 4.185 | 7.209 | -0.827 | -2.708 | -5.529 |
| | 5 | 1.031 | 1.449 | 1.613 | 1.856 | 4.190 | 7.217 | -0.825 | -2.741 | -5.604 |
| Asymmetric Trimming | | | | | | | | | | |
| $q = 0$ | 1 | 1.024 | 1.518 | 1.796 | 1.853 | 4.179 | 7.198 | -0.829 | -2.661 | -5.402 |
| | 3 | 1.016 | 1.635 | 2.096 | 1.845 | 4.147 | 7.137 | -0.829 | -2.512 | -5.041 |
| | 5 | 1.012 | 1.688 | 2.227 | 1.841 | 4.134 | 7.111 | -0.829 | -2.446 | -4.884 |
| $q = 0.01$ | 1 | 1.026 | 1.489 | 1.718 | 1.855 | 4.186 | 7.212 | -0.829 | -2.697 | -5.494 |
| | 3 | 1.022 | 1.531 | 1.815 | 1.852 | 4.175 | 7.190 | -0.830 | -2.644 | -5.375 |
| | 5 | 1.023 | 1.514 | 1.769 | 1.853 | 4.177 | 7.194 | -0.830 | -2.663 | -5.425 |
| $q = 0.05$ | 1 | 1.030 | 1.436 | 1.591 | 1.857 | 4.196 | 7.231 | -0.828 | -2.760 | -5.640 |
| | 3 | 1.030 | 1.428 | 1.572 | 1.857 | 4.195 | 7.229 | -0.827 | -2.767 | -5.656 |
| | 5 | 1.033 | 1.396 | 1.504 | 1.858 | 4.200 | 7.238 | -0.825 | -2.804 | -5.733 |
| Truncation | | | | | | | | | | |
| | 1 | 1.024 | 1.518 | 1.797 | 1.853 | 4.179 | 7.198 | -0.829 | -2.661 | -5.402 |
| | 3 | 1.016 | 1.636 | 2.099 | 1.845 | 4.147 | 7.136 | -0.829 | -2.511 | -5.038 |
| | 5 | 1.012 | 1.693 | 2.245 | 1.841 | 4.132 | 7.107 | -0.829 | -2.438 | -4.862 |

[a] Values of $\mu_w^{(1)}(L)$, $\mu_w^{(0)}(L)$ and $\Delta_w(L)$ are evaluated using a sufficiently large sample with 1000,000 observations.

**Table 2.** Bias of $\hat{\mu}_w^{(1)}(L)$, $\hat{\mu}_w^{(0)}(L)$ and $\hat{\Delta}_w(L)$ in the Presence of Increasingly Strong Tails in the Propensity Score Distribution (Simulation Study).

| Estimator | $\gamma$ | $\hat{\mu}_w^{(1)}(L)$ | | | $\hat{\mu}_w^{(0)}(L)$ | | | $\hat{\Delta}_w(L)$ | | |
|---|---|---|---|---|---|---|---|---|---|---|
| | | $L=2$ | $L=5$ | $L=10$ | $L=2$ | $L=5$ | $L=10$ | $L=2$ | $L=5$ | $L=10$ |
| Overlap Weighting | | | | | | | | | | |
| | 1 | 0.33 | 0.76 | 1.55 | -0.01 | 0.01 | -0.08 | -0.43 | -0.41 | -0.60 |
| | 3 | 0.20 | 0.69 | 2.06 | -0.03 | 0.06 | 0.12 | -0.30 | -0.30 | -0.53 |
| | 5 | 0.19 | 0.71 | 2.73 | -0.02 | 0.09 | 0.12 | -0.27 | -0.26 | -0.72 |
| Inverse Probability Treatment Weighting | | | | | | | | | | |
| No Trimming | | | | | | | | | | |
| | 1 | 0.33 | 0.78 | 1.58 | -0.01 | -0.01 | -0.16 | -0.43 | -0.46 | -0.74 |
| | 3 | 0.13 | 0.48 | 1.50 | -0.05 | -0.08 | -0.39 | -0.28 | -0.45 | -1.17 |
| | 5 | -2.48 | -3.40 | -2.88 | -0.21 | -0.46 | -1.31 | 2.56 | 1.57 | -0.59 |
| Symmetric Trimming | | | | | | | | | | |
| $a = 0.05$ | 1 | 0.33 | 0.78 | 1.58 | -0.01 | -0.01 | -0.16 | -0.43 | -0.46 | -0.74 |
| | 3 | 0.23 | 0.61 | 1.83 | -0.04 | 0.02 | 0.04 | -0.37 | -0.33 | -0.61 |
| | 5 | 0.25 | 0.78 | 3.14 | 0.02 | 0.09 | 0.09 | -0.26 | -0.31 | -0.97 |
| $a = 0.1$ | 1 | 0.34 | 0.73 | 1.46 | -0.01 | 0.00 | -0.15 | -0.43 | -0.42 | -0.68 |
| | 3 | 0.21 | 0.63 | 2.25 | -0.04 | 0.02 | 0.07 | -0.34 | -0.33 | -0.64 |
| | 5 | 0.12 | 0.64 | 3.25 | 0.00 | 0.13 | 0.20 | -0.15 | -0.15 | -0.74 |
| $a = 0.15$ | 1 | 0.37 | 0.65 | 1.23 | 0.00 | 0.01 | -0.13 | -0.46 | -0.36 | -0.58 |
| | 3 | 0.19 | 0.80 | 2.94 | -0.03 | 0.08 | 0.16 | -0.31 | -0.32 | -0.69 |
| | 5 | 0.34 | 1.06 | 4.13 | -0.01 | 0.13 | 0.23 | -0.46 | -0.37 | -0.90 |
| Asymmetric Trimming | | | | | | | | | | |
| $q = 0$ | 1 | 0.42 | 0.55 | 1.06 | 0.03 | 0.07 | -0.05 | -0.45 | -0.21 | -0.42 |
| | 3 | 2.60 | 2.46 | 3.04 | 0.27 | 0.57 | 0.65 | -2.58 | -0.66 | -0.35 |
| | 5 | 5.09 | 4.28 | 4.87 | 0.65 | 1.35 | 1.58 | -4.77 | -0.67 | 0.08 |
| $q = 0.01$ | 1 | 0.41 | 0.75 | 1.52 | 0.00 | 0.03 | -0.06 | -0.50 | -0.38 | -0.55 |
| | 3 | 0.56 | 0.93 | 2.44 | -0.01 | 0.04 | 0.10 | -0.72 | -0.47 | -0.68 |
| | 5 | 0.35 | 0.79 | 3.32 | 0.07 | 0.17 | 0.23 | -0.29 | -0.18 | -0.78 |
| $q = 0.05$ | 1 | 0.37 | 1.02 | 2.37 | -0.04 | 0.00 | -0.01 | -0.55 | -0.53 | -0.67 |
| | 3 | 0.13 | 0.84 | 3.24 | 0.01 | 0.13 | 0.25 | -0.14 | -0.24 | -0.58 |
| | 5 | 0.36 | 1.49 | 5.75 | -0.10 | 0.13 | 0.31 | -0.66 | -0.55 | -1.12 |
| Truncation | | | | | | | | | | |
| $a = 0.025$ | 1 | -0.10 | -0.03 | 0.31 | -0.05 | -0.08 | -0.27 | 0.02 | -0.11 | -0.46 |
| | 3 | -0.77 | -1.19 | -1.04 | -0.11 | -0.19 | -0.53 | 0.70 | 0.47 | -0.32 |
| | 5 | -2.68 | -3.81 | -3.50 | -0.19 | -0.45 | -1.30 | 2.85 | 1.88 | -0.29 |
| $a = 0.05$ | 1 | -0.59 | -0.88 | -0.92 | -0.08 | -0.16 | -0.40 | 0.55 | 0.25 | -0.22 |
| | 3 | -1.81 | -3.03 | -3.72 | -0.18 | -0.33 | -0.75 | 1.82 | 1.42 | 0.49 |
| | 5 | -3.38 | -5.08 | -5.33 | -0.23 | -0.53 | -1.41 | 3.62 | 2.63 | 0.41 |
| $a = 0.1$ | 1 | -1.67 | -2.64 | -3.27 | -0.16 | -0.33 | -0.68 | 1.70 | 0.98 | 0.18 |
| | 3 | -4.13 | -6.92 | -9.09 | -0.35 | -0.71 | -1.33 | 4.28 | 3.34 | 1.91 |
| | 5 | -5.76 | -9.09 | -10.96 | -0.41 | -0.91 | -1.95 | 6.11 | 4.76 | 2.21 |

**Table 3.** Relative Efficiency of $\hat{\mu}_w^{(1)}(L)$, $\hat{\mu}_w^{(0)}(L)$ and $\hat{\Delta}_w(L)$ Relative to $\hat{\mu}_{IPTW}^{(1)}(L)$, $\hat{\mu}_{IPTW}^{(0)}(L)$ and $\hat{\Delta}_{IPTW}(L)$ in the Presence of Increasingly Strong Tails in the Propensity Score Distribution.

| Estimator | $\gamma$ | $\hat{\mu}_w^{(1)}(L)$ | | | $\hat{\mu}_w^{(0)}(L)$ | | | $\hat{\Delta}_w(L)$ | | |
|---|---|---|---|---|---|---|---|---|---|---|
| | | $L=2$ | $L=5$ | $L=10$ | $L=2$ | $L=5$ | $L=10$ | $L=2$ | $L=5$ | $L=10$ |
| Overlap Weighting | | | | | | | | | | |
| | 1 | 1.00 | 1.08 | 1.23 | 1.00 | 1.05 | 1.07 | 1.00 | 1.08 | 1.10 |
| | 3 | 2.21 | 3.60 | 5.49 | 1.50 | 1.72 | 1.70 | 2.13 | 2.97 | 3.13 |
| | 5 | 5.72 | 9.71 | 14.79 | 3.36 | 3.05 | 2.65 | 5.42 | 7.02 | 6.83 |
| Inverse Probability Treatment Weighting | | | | | | | | | | |
| No Trimming | | | | | | | | | | |
| | 1 | 1.00 | 1.00 | 1.00 | 1.00 | 1.00 | 1.00 | 1.00 | 1.00 | 1.00 |
| | 3 | 1.00 | 1.00 | 1.00 | 1.00 | 1.00 | 1.00 | 1.00 | 1.00 | 1.00 |
| | 5 | 1.00 | 1.00 | 1.00 | 1.00 | 1.00 | 1.00 | 1.00 | 1.00 | 1.00 |
| Symmetric Trimming | | | | | | | | | | |
| $a = 0.05$ | 1 | 1.00 | 1.00 | 1.00 | 1.00 | 1.00 | 1.00 | 1.00 | 1.00 | 1.00 |
| | 3 | 2.03 | 2.58 | 3.15 | 1.29 | 1.35 | 1.32 | 1.91 | 2.18 | 2.16 |
| | 5 | 4.56 | 6.35 | 7.91 | 2.93 | 2.52 | 2.07 | 4.43 | 4.93 | 4.56 |
| $a = 0.1$ | 1 | 1.00 | 1.02 | 1.03 | 1.00 | 1.00 | 1.00 | 1.00 | 1.01 | 1.00 |
| | 3 | 1.94 | 2.91 | 4.21 | 1.27 | 1.45 | 1.48 | 1.88 | 2.51 | 2.66 |
| | 5 | 4.62 | 7.43 | 10.98 | 2.59 | 2.36 | 2.06 | 4.16 | 5.10 | 5.01 |
| $a = 0.15$ | 1 | 1.01 | 1.04 | 1.09 | 0.98 | 0.99 | 0.99 | 1.00 | 1.03 | 1.02 |
| | 3 | 1.64 | 2.74 | 4.62 | 1.20 | 1.47 | 1.49 | 1.59 | 2.29 | 2.63 |
| | 5 | 4.31 | 7.61 | 12.49 | 2.25 | 2.15 | 1.94 | 3.87 | 5.20 | 5.12 |
| Asymmetric Trimming | | | | | | | | | | |
| $q = 0$ | 1 | 0.98 | 1.00 | 1.02 | 1.00 | 0.99 | 1.00 | 0.98 | 0.99 | 0.99 |
| | 3 | 0.97 | 1.06 | 1.10 | 0.95 | 0.95 | 0.95 | 0.91 | 0.96 | 0.98 |
| | 5 | 0.92 | 0.94 | 0.94 | 0.92 | 0.92 | 0.91 | 0.86 | 0.86 | 0.85 |
| $q = 0.01$ | 1 | 0.87 | 0.92 | 1.06 | 0.96 | 1.03 | 1.05 | 0.89 | 0.98 | 1.04 |
| | 3 | 1.40 | 2.04 | 2.95 | 1.20 | 1.29 | 1.31 | 1.35 | 1.73 | 1.97 |
| | 5 | 2.78 | 4.48 | 6.47 | 2.69 | 2.36 | 1.91 | 2.69 | 3.49 | 3.66 |
| $q = 0.05$ | 1 | 0.68 | 0.77 | 1.03 | 0.80 | 0.91 | 0.99 | 0.71 | 0.86 | 0.99 |
| | 3 | 1.20 | 2.28 | 4.30 | 0.99 | 1.14 | 1.21 | 1.17 | 1.77 | 2.12 |
| | 5 | 2.75 | 5.72 | 11.24 | 1.68 | 1.53 | 1.46 | 2.57 | 3.83 | 3.91 |
| Truncation | | | | | | | | | | |
| $a = 0.025$ | 1 | 1.01 | 1.04 | 1.09 | 1.00 | 1.02 | 1.02 | 1.01 | 1.03 | 1.04 |
| | 3 | 1.43 | 1.53 | 1.63 | 1.14 | 1.14 | 1.10 | 1.37 | 1.39 | 1.37 |
| | 5 | 1.15 | 1.18 | 1.17 | 1.29 | 1.10 | 1.04 | 1.17 | 1.16 | 1.13 |
| $a = 0.05$ | 1 | 1.02 | 1.07 | 1.15 | 1.00 | 1.04 | 1.05 | 1.02 | 1.06 | 1.07 |
| | 3 | 1.73 | 1.94 | 2.15 | 1.30 | 1.30 | 1.22 | 1.64 | 1.71 | 1.65 |
| | 5 | 1.54 | 1.64 | 1.64 | 1.52 | 1.24 | 1.13 | 1.53 | 1.53 | 1.45 |
| $a = 0.1$ | 1 | 1.03 | 1.12 | 1.25 | 1.01 | 1.06 | 1.09 | 1.03 | 1.10 | 1.13 |
| | 3 | 2.21 | 2.69 | 3.20 | 1.48 | 1.55 | 1.46 | 2.06 | 2.28 | 2.22 |
| | 5 | 2.84 | 3.23 | 3.42 | 2.14 | 1.76 | 1.49 | 2.74 | 2.76 | 2.55 |

**Table 4.** Coverage Rates (%) of the 95% Confidence Intervals for Estimators $\hat{\mu}_w^{(1)}(L)$, $\hat{\mu}_w^{(0)}(L)$ and $\hat{\Delta}_w(L)$ in the Presence of Increasingly Strong Tails in the Propensity Score Distribution.

| Estimator | $\gamma$ | $\hat{\mu}_w^{(1)}(L)$ | | | $\hat{\mu}_w^{(0)}(L)$ | | | $\hat{\Delta}_w(L)$ | | |
|---|---|---|---|---|---|---|---|---|---|---|
| | | $L=2$ | $L=5$ | $L=10$ | $L=2$ | $L=5$ | $L=10$ | $L=2$ | $L=5$ | $L=10$ |
| Overlap Weighting | | | | | | | | | | |
| | 1 | 96.8 | 97.6 | 97.6 | 94.4 | 94.6 | 93.0 | 96.7 | 96.5 | 94.9 |
| | 3 | 96.6 | 96.4 | 96.3 | 94.8 | 94.5 | 94.7 | 96.0 | 96.1 | 95.7 |
| | 5 | 95.4 | 96.4 | 95.5 | 92.8 | 95.3 | 93.6 | 94.7 | 95.0 | 94.8 |
| Inverse Probability Treatment Weighting | | | | | | | | | | |
| No Trimming | | | | | | | | | | |
| | 1 | 97.3 | 97.5 | 97.6 | 94.9 | 94.3 | 92.2 | 96.4 | 96.3 | 94.6 |
| | 3 | 94.5 | 91.8 | 87.1 | 93.0 | 92.4 | 90.6 | 92.6 | 90.3 | 86.7 |
| | 5 | 80.1 | 75.3 | 70.1 | 87.8 | 87.3 | 84.9 | 77.6 | 73.5 | 74.5 |
| Symmetric Trimming | | | | | | | | | | |
| $a=0.05$ | 1 | 97.2 | 97.5 | 97.6 | 95.0 | 94.4 | 92.2 | 96.4 | 96.3 | 94.6 |
| | 3 | 97.0 | 96.2 | 95.1 | 93.1 | 93.0 | 92.1 | 96.6 | 94.8 | 93.4 |
| | 5 | 95.8 | 95.7 | 94.9 | 91.8 | 93.6 | 92.6 | 95.2 | 94.4 | 92.5 |
| $a=0.1$ | 1 | 97.2 | 97.6 | 97.8 | 94.9 | 94.4 | 92.3 | 96.5 | 96.5 | 94.6 |
| | 3 | 96.3 | 96.4 | 95.2 | 93.3 | 93.6 | 94.2 | 95.8 | 95.4 | 95.6 |
| | 5 | 95.1 | 94.2 | 93.4 | 91.8 | 93.4 | 92.0 | 94.8 | 93.1 | 92.9 |
| $a=0.15$ | 1 | 97.1 | 97.8 | 97.5 | 94.8 | 94.5 | 92.3 | 96.9 | 97.4 | 94.6 |
| | 3 | 94.3 | 94.7 | 94.0 | 94.4 | 95.0 | 94.8 | 94.9 | 94.7 | 94.6 |
| | 5 | 94.7 | 94.6 | 94.9 | 92.3 | 94.4 | 92.1 | 94.3 | 93.8 | 93.4 |
| Asymmetric Trimming | | | | | | | | | | |
| $q=0$ | 1 | 97.1 | 97.6 | 97.5 | 94.7 | 94.2 | 92.6 | 96.2 | 96.3 | 95.0 |
| | 3 | 91.2 | 92.4 | 88.0 | 89.0 | 89.6 | 89.3 | 88.7 | 88.0 | 84.1 |
| | 5 | 81.5 | 80.9 | 76.6 | 77.1 | 78.5 | 81.4 | 77.9 | 72.0 | 69.7 |
| $q=0.01$ | 1 | 95.6 | 96.2 | 96.4 | 94.7 | 94.3 | 92.7 | 94.9 | 95.7 | 94.9 |
| | 3 | 92.0 | 93.5 | 91.8 | 92.6 | 91.4 | 92.7 | 91.5 | 90.7 | 90.6 |
| | 5 | 88.4 | 89.3 | 89.4 | 91.9 | 93.0 | 90.6 | 88.8 | 87.6 | 88.4 |
| $q=0.05$ | 1 | 94.4 | 94.3 | 94.7 | 94.4 | 94.0 | 93.5 | 94.2 | 93.6 | 94.3 |
| | 3 | 91.7 | 91.8 | 92.7 | 93.9 | 94.2 | 94.0 | 91.4 | 91.0 | 91.7 |
| | 5 | 90.1 | 91.2 | 93.4 | 91.2 | 92.6 | 92.4 | 91.2 | 90.9 | 92.2 |
| Truncation | | | | | | | | | | |
| $a=0.025$ | 1 | 97.1 | 96.9 | 97.8 | 94.8 | 94.9 | 92.5 | 96.4 | 96.5 | 94.8 |
| | 3 | 95.4 | 92.3 | 88.3 | 93.7 | 93.1 | 91.3 | 94.1 | 92.0 | 89.4 |
| | 5 | 80.6 | 75.8 | 70.5 | 87.9 | 87.4 | 84.9 | 78.1 | 73.8 | 74.6 |
| $a=0.05$ | 1 | 96.8 | 96.5 | 96.7 | 94.8 | 95.0 | 92.6 | 96.7 | 96.3 | 94.6 |
| | 3 | 94.6 | 92.1 | 88.4 | 94.3 | 93.8 | 91.2 | 94.0 | 92.5 | 90.3 |
| | 5 | 82.6 | 77.7 | 72.9 | 89.4 | 89.1 | 85.8 | 79.9 | 76.7 | 77.3 |
| $a=0.1$ | 1 | 94.1 | 93.8 | 94.4 | 95.0 | 95.2 | 93.0 | 95.3 | 95.9 | 95.3 |
| | 3 | 90.2 | 86.3 | 82.1 | 95.2 | 94.2 | 91.5 | 91.9 | 90.8 | 91.6 |
| | 5 | 83.2 | 77.9 | 72.7 | 93.0 | 92.8 | 88.5 | 83.2 | 81.0 | 83.1 |

**Table 5.** Bias, Relative Efficiency and Coverage Rates (%) of the 95% Confidence Intervals for Estimators of $\hat{\mu}_w^{(1)}(L)$, $\hat{\mu}_w^{(0)}(L)$ and $\widehat{\Delta}_w(L)$ in the Presence of Increasingly Strong Tails in the Propensity Score Distribution (Simulation Study)[b] with Both Closed-form Variance and Bootstrap Variance under Sample Size $n = 250$.

| | $\gamma$ | $\hat{\mu}_w^{(1)}(L)$ | | | $\hat{\mu}_w^{(0)}(L)$ | | | $\widehat{\Delta}_w(L)$ | | |
|---|---|---|---|---|---|---|---|---|---|---|
| | | $L=2$ | $L=5$ | $L=10$ | $L=2$ | $L=5$ | $L=10$ | $L=2$ | $L=5$ | $L=10$ |
| **Bias** | | Overlap Weighting | | | | | | | | |
| | 1 | 0.00 | 0.37 | 1.83 | 0.02 | 0.15 | 0.37 | 0.11 | -0.39 | -1.66 |
| | 3 | 0.09 | 0.52 | 2.18 | 0.17 | 0.33 | 0.58 | 0.63 | -0.15 | -1.57 |
| | 5 | 0.16 | 0.89 | 3.10 | 0.20 | 0.36 | 0.76 | 0.43 | -0.98 | -2.33 |
| | | Inverse Probability Treatment Weighting | | | | | | | | |
| | 1 | -0.02 | 0.38 | 1.91 | 0.01 | 0.10 | 0.21 | 0.12 | -0.57 | -2.19 |
| | 3 | -0.41 | -0.46 | 0.58 | 0.08 | 0.11 | -0.18 | 2.73 | 1.48 | -1.25 |
| | 5 | -1.51 | -2.85 | -2.92 | -0.22 | -0.37 | -0.80 | 6.75 | 5.60 | 2.47 |
| **Relative Efficiency** | | Overlap Weighting | | | | | | | | |
| | 1 | 1.00 | 1.00 | 1.03 | 1.05 | 1.05 | 1.05 | 1.00 | 1.02 | 1.05 |
| | 3 | 1.60 | 1.66 | 1.91 | 1.64 | 1.52 | 1.38 | 1.64 | 1.61 | 1.65 |
| | 5 | 1.95 | 2.14 | 2.58 | 2.02 | 1.80 | 1.49 | 2.00 | 1.93 | 1.92 |
| | | Inverse Probability Treatment Weighting | | | | | | | | |
| | 1 | 1.00 | 1.00 | 1.00 | 1.00 | 1.00 | 1.00 | 1.00 | 1.00 | 1.00 |
| | 3 | 1.00 | 1.00 | 1.00 | 1.00 | 1.00 | 1.00 | 1.00 | 1.00 | 1.00 |
| | 5 | 1.00 | 1.00 | 1.00 | 1.00 | 1.00 | 1.00 | 1.00 | 1.00 | 1.00 |
| **Coverage Rates (Sandwich variance)** | | Overlap Weighting | | | | | | | | |
| | 1 | 94.6 | 95.3 | 96.1 | 93.6 | 94.5 | 93.1 | 95.5 | 94.8 | 94.2 |
| | 3 | 93.7 | 92.1 | 91.3 | 90.8 | 92.3 | 91.6 | 94.6 | 93.1 | 90.8 |
| | 5 | 92.4 | 92.3 | 92.3 | 89.2 | 91.5 | 90.9 | 93.8 | 92.1 | 91.5 |
| | | Inverse Probability Treatment Weighting | | | | | | | | |
| | 1 | 96.3 | 96.3 | 96.1 | 93.6 | 93.2 | 92.8 | 95.5 | 94.4 | 94.5 |
| | 3 | 93.7 | 91.6 | 88.3 | 87.0 | 88.1 | 87.4 | 91.4 | 88.6 | 86.5 |
| | 5 | 90.2 | 84.5 | 79.3 | 84.9 | 86.2 | 84.7 | 87.1 | 83.9 | 84.7 |
| **Coverage Rates (Bootstrap)** | | Overlap Weighting | | | | | | | | |
| | 1 | 94.6 | 95.2 | 94.6 | 93.6 | 94.5 | 93.9 | 95.7 | 94.4 | 94.2 |
| | 3 | 93.9 | 92.6 | 91.0 | 91.8 | 92.8 | 93.0 | 95.1 | 93.1 | 91.5 |
| | 5 | 93.5 | 93.7 | 92.2 | 90.4 | 92.7 | 93.2 | 95.5 | 92.9 | 92.7 |
| | | Inverse Probability Treatment Weighting | | | | | | | | |
| | 1 | 95.4 | 95.5 | 94.8 | 93.6 | 93.7 | 93.3 | 95.4 | 94.4 | 94.9 |
| | 3 | 94.0 | 91.4 | 88.3 | 88.6 | 90.2 | 90.6 | 93.7 | 90.1 | 89.3 |
| | 5 | 91.6 | 86.7 | 81.9 | 88.7 | 91.2 | 89.9 | 90.9 | 88.3 | 88.5 |

[b]In this simulation, PS term is omitted in generating the potential survival outcomes in the simulation design and true values of $\mu_w^{(1)}(L), \mu_w^{(0)}(L)$ and $\Delta_w(L)$ are the same as those in Web Table 5.